\documentclass{kapproc}
\usepackage{procps} 
\usepackage[dvips]{graphicx}
\upperandlowercase
\kluwerbib

\begin{document}
\articletitle{Status of ANITA and ANITA-lite}
\author{Andrea Silvestri$^{\dagger}$ for the ANITA Collaboration$^{\ast}$}
\affil{$^{\dagger}$Department of Physics and Astronomy, University of
California, Irvine, CA 92697, USA}
\email{silvestri@HEP.ps.uci.edu}
\vskip0.1in
{\footnotesize
\noindent
{$^{\ast}$The ANITA Collaboration:}\\
\noindent
S.~W.~Barwick\altaffilmark{3},
J.~J.~Beatty\altaffilmark{7}, 
D.~Z.~Besson\altaffilmark{6},
W.~R.~Binns\altaffilmark{8},
B.~Cai\altaffilmark{10},
J.~M.~Clem\altaffilmark{1},
A.~Connolly\altaffilmark{4},
D.~F.~Cowen\altaffilmark{2},
P.~F.~Dowkontt\altaffilmark{8},
M.~A.~DuVernois\altaffilmark{10}, 
P.~A.~Evenson\altaffilmark{1},
D.~Goldstein\altaffilmark{3},
P.~W.~Gorham\altaffilmark{5},
C.~L.~Hebert\altaffilmark{5},
M.~H.~Israel\altaffilmark{8},
H.~Krawzczynski\altaffilmark{8},
J.~G.~Learned\altaffilmark{5},
K.~M.~Liewer\altaffilmark{9},
J.~T.~Link\altaffilmark{5},
S.~Matsuno\altaffilmark{5},
P.~Miocinovic\altaffilmark{5},
J.~Nam\altaffilmark{3},
C.~J.~Naudet\altaffilmark{9},
R.~Nichol\altaffilmark{7},
M.~Rosen\altaffilmark{5},
D.~Saltzberg\altaffilmark{4},
D.~Seckel\altaffilmark{1},
A.~Silvestri\altaffilmark{3},
G.~S.~Varner\altaffilmark{5},
F.~Wu\altaffilmark{3}.
\vspace{2.mm}
\noindent
\affil
{\altaffilmark{1}{Bartol Research Institute, University of Delaware,
Newark, DE 19716, USA}, \
\altaffilmark{2}{Dept. of Astronomy and Astrophysics, Penn. State
University, University Park, PA 16802, USA}, \
\altaffilmark{3}{Dept. of Physics and Astronomy, University of
California, Irvine CA 92697, USA}, \
\altaffilmark{4}{Dept. of Physics and Astronomy, University of
California, Los Angeles, CA 90095, USA}, \
\altaffilmark{5}{Dept. of Physics and Astronomy, University of
Hawaii, Manoa, HI 96822, USA}, \
\altaffilmark{6}{Dept. of Physics and Astronomy, University of
Kansas, Lawrence, KS 66045, USA}, \
\altaffilmark{7}{Dept. of Physics, Ohio State University, Columbus,
OH 43210, USA}, \
\altaffilmark{8}{Dept. of Physics, Washington University in
St. Louis, MO 63130, USA}, \
\altaffilmark{9}{Jet Propulsion Laboratory, Pasadena, CA 91109, USA}, \
\altaffilmark{10}{School of Physics and Astronomy, University of
Minnesota, Minneapolis, MN 55455, USA} \
}
}
\begin{abstract}
We describe a new experiment to search for neutrinos with energies above
$3 \times 10^{18}$ eV based on the observation of short duration radio pulses 
that are emitted from neutrino-initiated cascades.  
The primary objective of the ANtarctic Impulse Transient Antenna
(ANITA) mission is to measure the flux of
Greisen-Zatsepin-Kuzmin (GZK) neutrinos and search for neutrinos
from Active Galactic Nuclei (AGN).
We present first results obtained from the successful launch of
a 2-antenna prototype instrument (called ANITA-lite) that circled 
Antarctica for 18 days during the 03/04 Antarctic campaign and show
preliminary results from attenuation length studies of electromagnetic
waves at radio frequencies in Antarctic ice. The ANITA detector is
funded by NASA, and the first flight is scheduled for December 2006.
\end{abstract}


\section*{Introduction}
ANITA is designed to search for particles that are created in the most powerful
environments in the universe.
Its primary goal is to discover neutrinos generated by
interactions between the highest energy cosmic rays and the cosmic
microwave background radiation, as well as neutrinos created at the
inner edge of supermassive black holes (SMBH).
Flying at an altitude between 35-40 km above the Antarctic continent, the
ANITA balloon-borne telescope will be sensitive to a target volume of
$\sim 10^{6}$ km$^{3}$ of radio-transparent ice. 
The aperture of ANITA exceeds $\sim$ 100 km$^{3} \cdot$sr at E$_{\nu} = 3
\times 10^{18}$ eV, averaged over neutrino flavor and assuming equal
fluxes of all flavors. The aperture continues to grow rapidly as
E$_{\nu}$ increases, reaching the order of $10^{5}$ km$^{3} \cdot$sr at
E$_{\nu} =  10^{21}$ eV.
At EeV energies, the Earth is
opaque to neutrinos so only horizontal or slightly downgoing neutrinos
are detectable (Fig.~\ref{fig1}). Neutrinos interact within the ice
and generate compact particle showers, which emit coherent radio
signals that are detectable in the radio-quiet environment of the
Antarctic continent.

\section{Science Goals}
\paragraph{Standard Sources}
High Energy neutrinos carry unique information from objects in the
universe, and complement the information provided by UHE $\gamma$-ray
astronomy. For example, neutrinos, unlike photons, can propagate
throughout the universe unattenuated, but photon astronomy between
10-100 TeV is limited to distances less than a few hundred Mpc due to
interactions with infrared background photons, and to even shorter
distances at PeV energies due to interactions with the cosmic
microwave background.
Neutrinos may be the only method to shed light on
acceleration processes associated with sources of the highest energy
cosmic rays, which extend more than six orders of magnitude above
energy of 100 TeV. Thus, the direct detection of high energy
neutrinos~\mbox{({\small{\cite{Stecker68,Berez69,stanev}}})}
will complement the investigation of the GZK
cutoff~({\small{\cite{Greisen_66,Zatsepin_66,kuzmin}}}),
one of the most controversial issues in cosmic ray physics.

Several theoretical models predict neutrino emission related to
accreting SMBH, which arise from particle acceleration
near the black hole at the center of AGN~({\small{\cite{Mannheim_01}}}).
In this environment ultra-high energy protons may escape from the
source, and a fraction of them eventually interact to produce high
energy neutrinos. Therefore, high-energy cosmic-ray
observations can be used to set a model-independent upper bound on the
high-energy neutrino flux~({\small{\cite{Waxman_99}}}).

\paragraph{Physics beyond the Standard Model}
The ANITA science goals extend beyond the Standard Model
if GZK neutrinos produce
highly unstable micro black holes (BH) when they interact with
ice~({\small{\cite{Feng02,Alv02}}}).
The decay of these highly unstable micro BH via Hawking radiation
would generate energetic hadronic showers that ANITA would
detect.
The signature of such events is the observation of an enhanced
detection rate that is strongly energy dependent.
This observation would provide evidence of new phenomena, such as the
existence of extra dimensions.

\section{ANITA}
\begin{figure}[ht]
\includegraphics[width=2.3 truein]{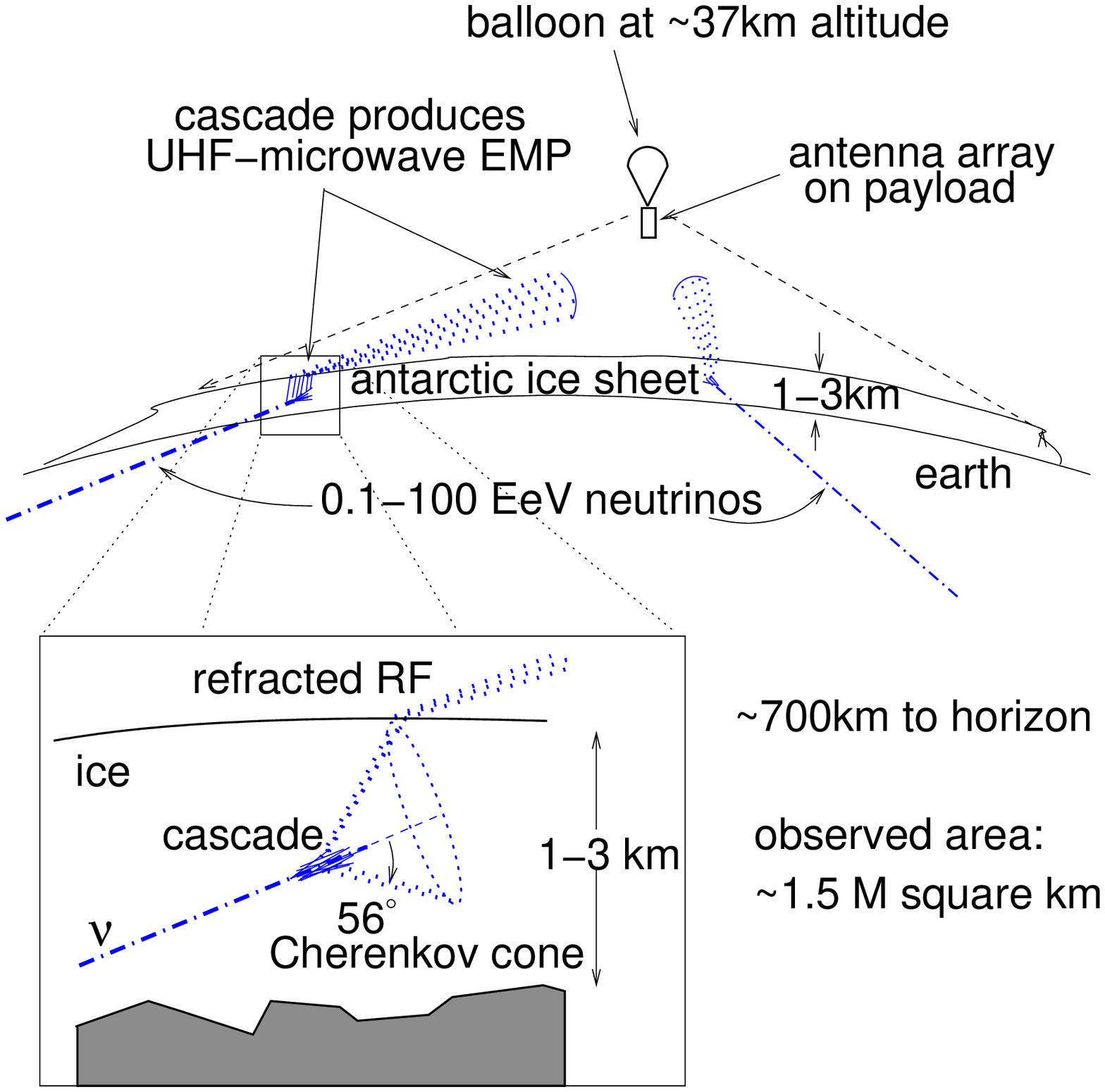}
~~~~~~~~\includegraphics[width=2.0 truein]{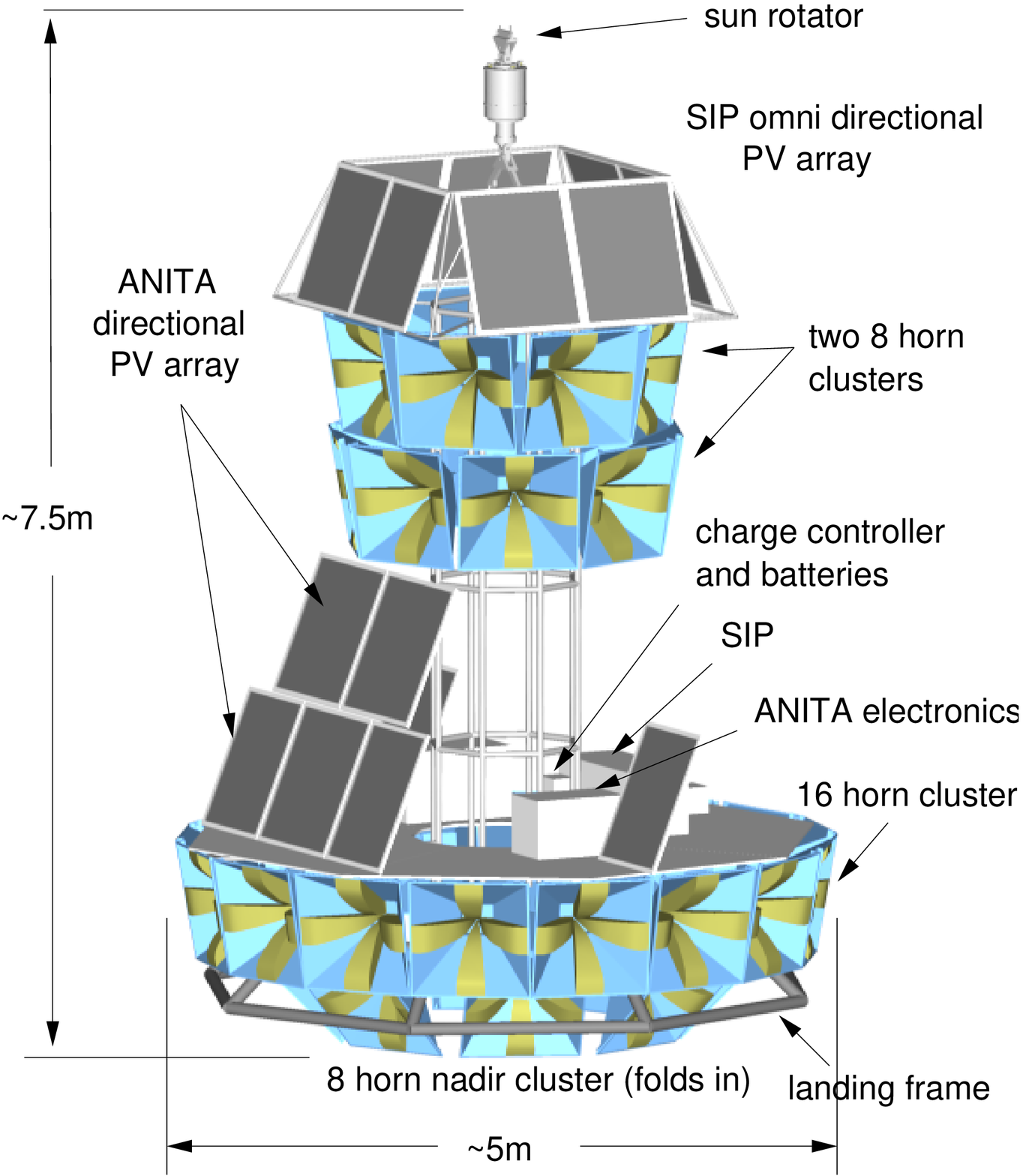}
\vskip-0.1in
\sidebyside
{\caption{\label{fig1} Schematic of the ANITA concept, displaying the basic
geometry for a detection of the coherent Cherenkov radio pulse generated
by the cascades in Antarctic ice.}}
{\caption{\label{fig2} Layout of the ANITA payload showing the geometry of
dual-polarization, quad-ridged antennas.}}
\end{figure}

\paragraph{The Concept}
The concept of detecting high energy particles through coherent
radio emission was first postulated by
Askaryan~({\small{\cite{Askaryan_62}}}) and has recently been
confirmed in accelerator experiments~({\small{\cite{Saltz01}}}).
Particle cascades induced by neutrinos in Antarctic ice are very
compact, no more than a few centimeters in lateral extent~({\small{\cite{Zas:1991jv}}}).
The resulting radio emission is coherent Cherenkov radiation
which is  characterized by a conical emission geometry,
broadband frequency content, and linear polarization. 
ANITA~({\small{\cite{barwick_1}}}) will observe the Antarctic ice sheet out to a horizon
approaching 700~km, monitoring a neutrino detection volume of order
$10^6$~km$^3$. 
The direction to the event is measured by time differences between
antennas in the upper and lower clusters (Fig.~\ref{fig2}).
The statistical distribution of events should correlate with ice
thickness, averaged over observable volume.
As illustrated in Fig.~\ref{fig1}, ANITA will search for radio
pulses that arise from electromagnetic and hadronic cascades within
the ice. The signals propagate through 1-3 km of ice with little
attenuation. 
At the energies of relevance to ANITA, the Earth strongly attenuates
the neutrino flux, so ANITA is primarily sensitive to horizontal
neutrinos.
A radio pulse with zenith angle $< 34^{\circ}$ will refract through
the air-ice interface and may be observed by ANITA.
Refraction and reflection effects due to small variations
of the index of refraction in the layered structure of Antarctic ice
are modest for the relevant incident angles.
The RF emission pattern is peaked in a forward conical geometry, but
considerable power remains within 5$^{\circ}$ of the Cherenkov
cone. 
At ANITA energies, cascades initiated by electrons (e.g. in $\nu_e$ charged
current events) are altered by
LPM~({\small{\cite{Landau:1953gr,Migdal:1957}}}) effects which narrows
the width of the Cerenkov Cone pattern to considerably less than
5$^{\circ}$. Cascades initiated by recoil hadrons are not affected, and
provide the bulk of observable events for neutrino energies above 1
EeV~({\small{\cite{Alvarez-Muniz:1998px}}}).

\paragraph{Effect of Non-uniform Snow Surface}
The ANITA team is investigating the effect of surface features on
signal characteristics. Surface structures that are somewhat larger
than the wavelength band of the Cherenkov pulse, such as sastrugi,
occupy only a small fraction of the surface. Random small-scale
features are not expected to produce significant variations from average
behavior. While the visual impact of surface roughness is quite
apparent, the density contrast (and gradient of index of refraction) is
only 10-30\%.
The broadband characteristics of the Cherenkov pulse also tend to
mitigate amplitude variation due to interference. 
An event located at a typical depth of 1 km produces a
coherent patch on the surface of $\sim$ 30 m in radius for the
frequencies of interest. 
We have performed an analytic treatment of the air interface, assuming Gaussian
fluctuations for the depth of snow, and find that the loss of phase
coherence at the antenna reduces the detected power by less than 5\%. We
are pursuing the use of optics analysis software to numerically study
surface effects.

\paragraph{The Detector}
The ANITA instrument~({\small{\cite{barwick_2}}}), shown in Fig.~\ref{fig2}, is a
cluster of broadband quad-ridged horn antennas with a frequency range
from 0.2 to 1.2 GHz.
The beam width of the antenna is about 60$^{\circ}$, with a gain of
approximately 9 dBi at 300MHz; gain is roughly constant across the entire
band. The detector geometry is defined by a
cylindrically symmetric array of 2 levels of 8 antennas each on the
upper portion, with a downward cant of about $10^{\circ}$, to achieve
complete coverage of the horizon down to within $40^{\circ}$ of the
nadir. The antenna beams overlap, giving redundant coverage in the
horizontal plane. 
A second array of 16 antennas on the lower portion of the
payload provides a vertical baseline for time delay measurements which will
determine pulse direction in elevation angle. The absolute azimuthal
orientation is established by Sun sensors, and payload tilt is
measured by differential GPS.

\section{ANITA-lite}
A two-antenna prototype of ANITA, called ANITA-lite, was flown for 18
days on a Long-Duration Balloon (LDB). It was launched on December 17,
2003 as a piggyback instrument onboard the Trans-Iron Galactic
Element Recorder (TIGER)~({\small{\cite{Link:2003mm}}}). Both
polarizations of each quad-ridged horn antenna were digitized. For the
purpose of triggering, the four linear polarization channels 
were converted to right-handed and left-handed circular polarization
for a total of four channels.
The trigger criteria required that 3 of 4 waveforms exceed a specified voltage
threshold, which varied throughout the flight.
The ANITA-lite mission tested nearly every subsystem of ANITA, and
monitored the Antarctic continent
for impulsive Radio Frequency Interference (RFI) and ambient thermal
noise levels as well as triggered events.
We discuss a few results that impact the upcoming ANITA experiment.

\begin{figure}[h]
 \includegraphics[width=2.3 truein]{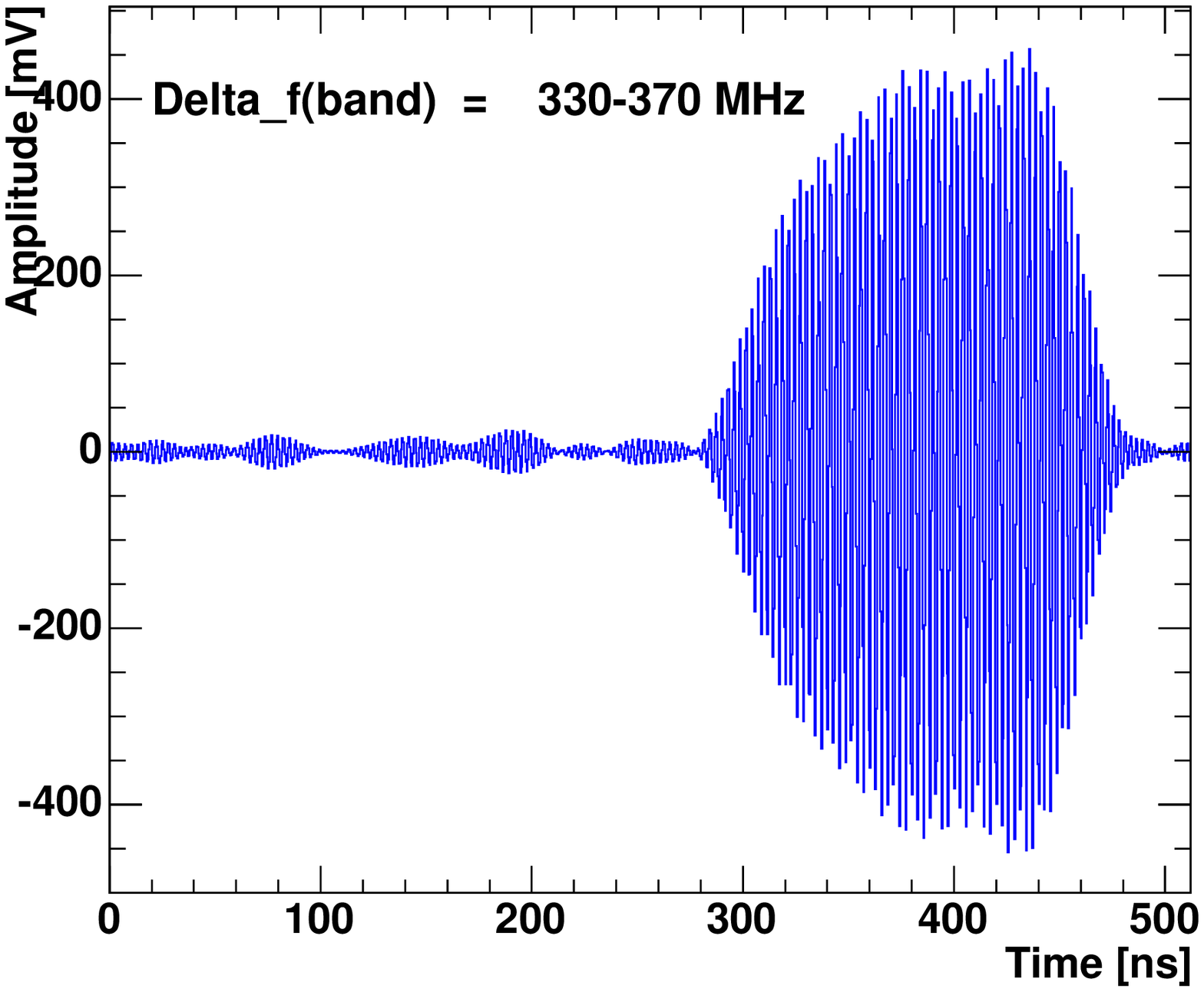}
~~~\includegraphics[width=2.3 truein]{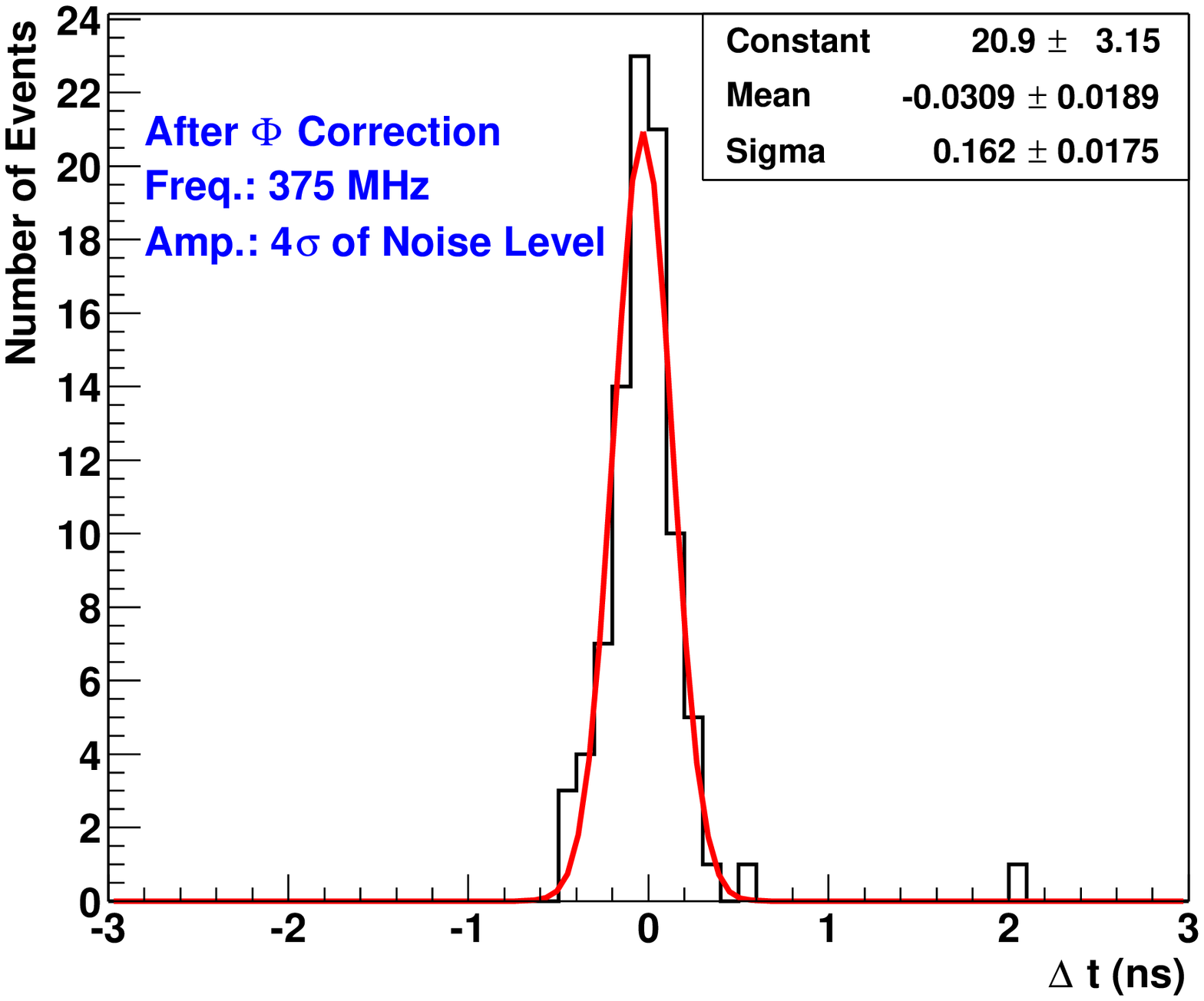}
\vskip-0.2in
\sidebyside
{\caption{\label{fig3} Calibration event at 350 MHz. Waveform
    after Band-Pass filter.}}
{\caption{\label{fig4} Distribution of time difference 
$\Delta t$ between copolarized receiver channels.}}
\end{figure}

\paragraph{Timing Analysis Results}
While aloft, {ANITA-lite} received signals from a surface transmitter
to measure the timing resolution. Signals were observed out to distances
of 200 km. 
The waveforms were processed by digital filtering of the frequency
components~({\small{\cite{andrea}}}), and the results of a band-pass
filter applied to the raw waveform can be seen in Fig.~\ref{fig3}.
The time reference for each antenna was extracted by interpolating
the zero-crossing of a signal that exceeds 4$V_{rms}$, where V$_{rms}$ is
the rms fluctuation of the noise voltage~\mbox{({\small{\cite{jiwoo}}})}.
The uncertainty in the time difference between the co-polarized
channels of both antennas flown on ANITA-lite was $\sigma_{\Delta t}=
0.16$~ns (see Fig.~\ref{fig4}) after correction for azimuthal variation.
This value is based on the full set of calibration events that
were acquired over a range of 40$^{\circ}$ in zenith angle $\theta$, indicating 
that the systematic dependence of timing on $\theta$ is rather weak.\par
For ANITA, we expect the time resolution between antenna clusters to
improve to $\sigma_{\Delta t} = 0.1$~ns due to the increase in the number of
measurements by the full array.
Using 3.3~m vertical separation between the upper and lower
antenna arrays, $d$, the expected intrinsic zenith angle resolution
$\sigma_{\theta}$ is
\begin{eqnarray}
{\sigma_{\theta} = (\sigma_{\Delta t} ~\frac{c}{d} ~57.3^{\circ}) = 0.5^{\circ}}
\label{eq:sigma}
\end{eqnarray}
for events
near the horizon.  Similarly, azimuthal angular resolution
$\sigma_{\phi}$ is estimated to be $1.5 ^{\circ}$ using the $\sim$
1 m baseline separation between antenna elements in a ring.

\paragraph{Thermal Noise Analysis Results}
Another goal of Anita-lite was to investigate the level of ambient
broadband noise in Antarctica at the balloon altitude and 
relate it to the absolute thermal noise background.
Cosmic background radiation combined with the Galactic radio noise
contributes noise of the order of 10-50K, and the average
temperature of the Antarctic ice surface is in the range of 220-250K. 
The contribution of the Sun at 0.2-1.2~GHz is between $10^5$ to $10^6$~K
depending on solar activity and radio frequency, however, its
angular size in the radio band is of order $1^{\circ}$ so its
contribution is comparable to the Galaxy when averaged over the
angular acceptance of the antenna.
Both the Sun and the brightest region of the Galaxy were within the same
field of view during the ANITA-lite flight.
\begin{figure*}[htb!]
\includegraphics[width=2.3 truein]{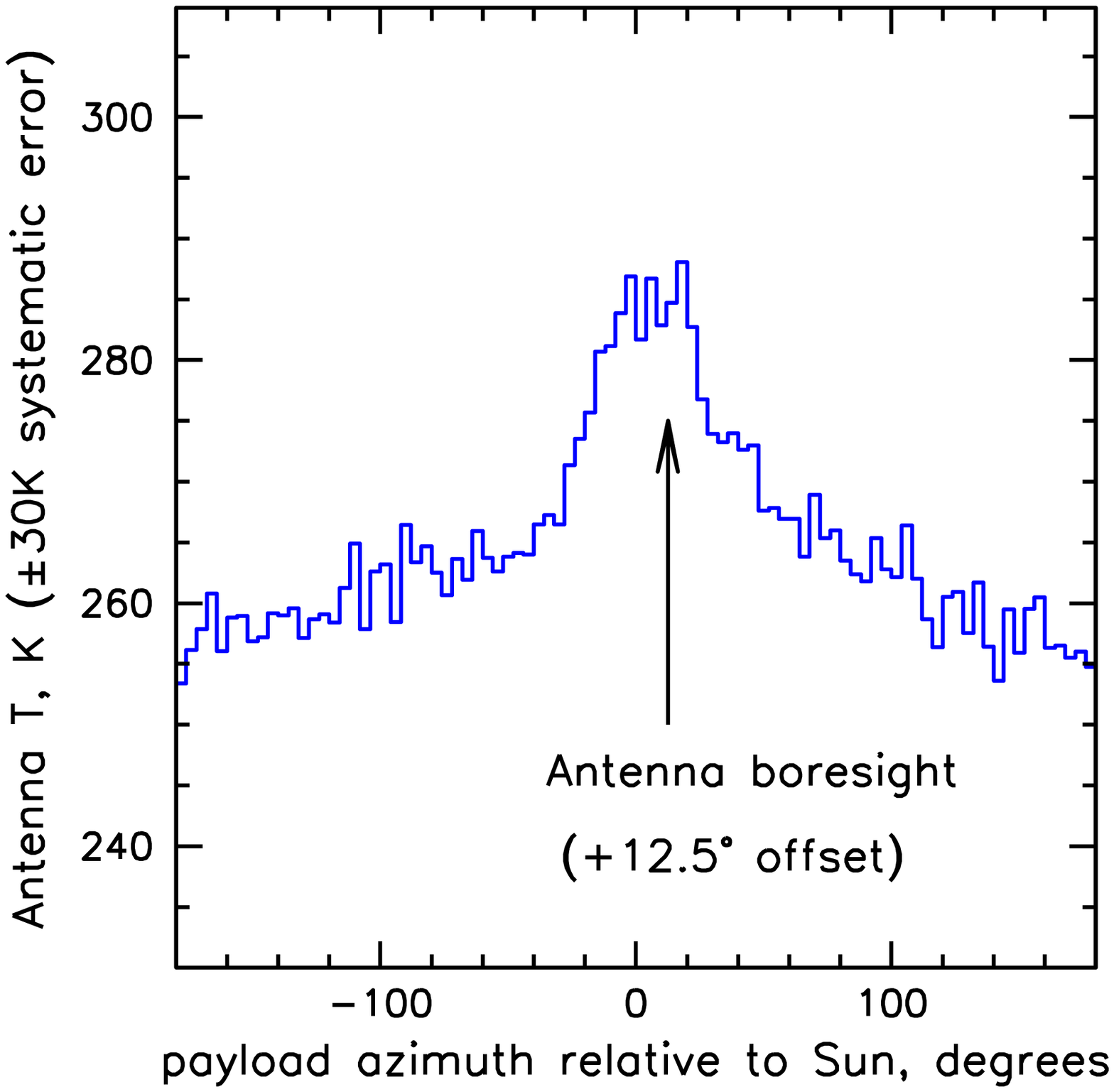}
~~~\includegraphics[width=2.3 truein]{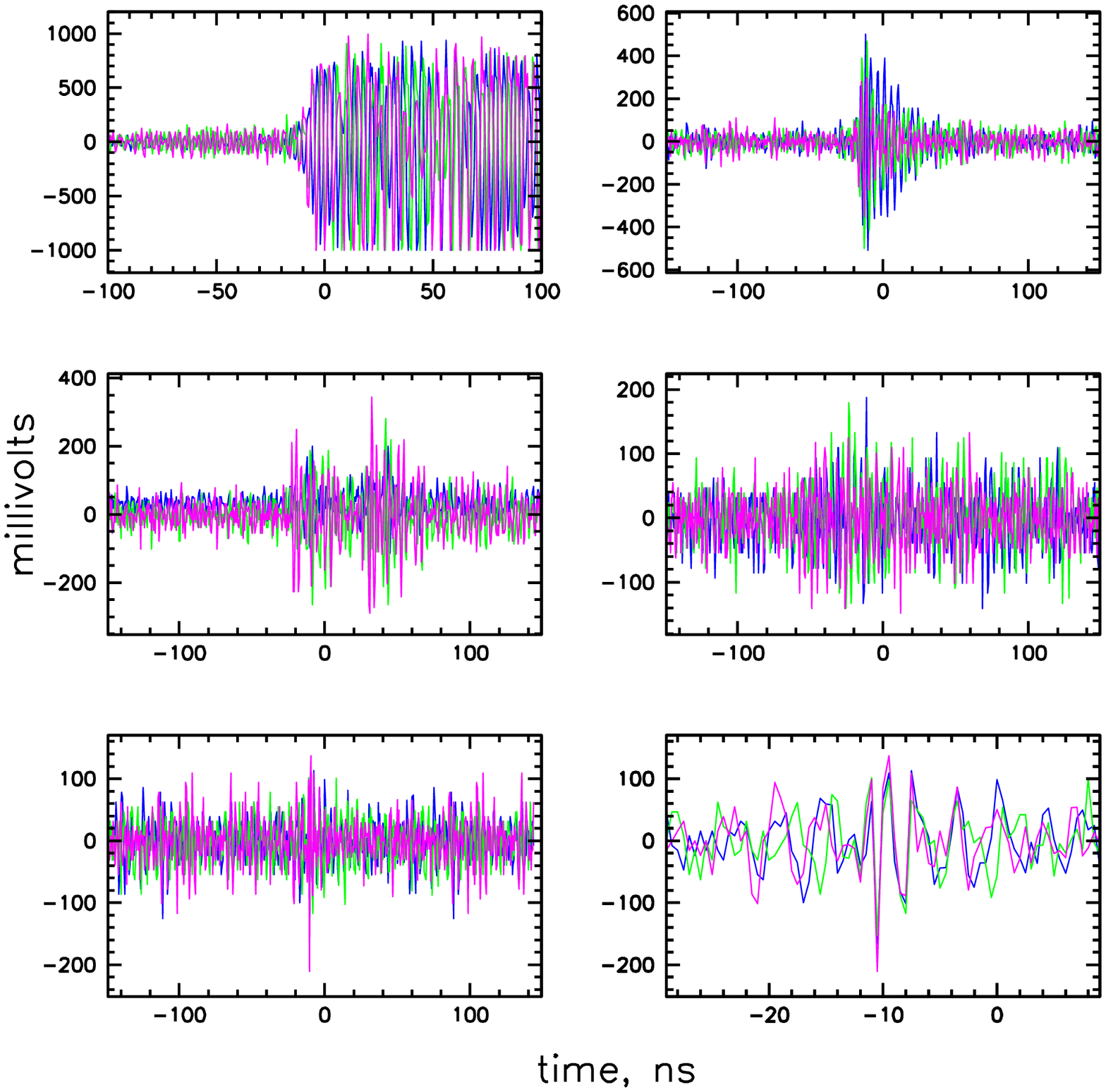}
\vskip-0.2in
\sidebyside
{\caption{\label{fig5} ANITA-lite calibration sources - Sun and Galaxy
    - thermal measurements as a function of angle with respect to the
    Sun during the 18 day flight.}}
{\caption{\label{fig6} Examples of ANITA-lite interference
events (top 4 panes) and two views of an injected 
signal-like impulse (bottom panes).}}
\end{figure*}
The results of this analysis are shown in Fig.~\ref{fig5}, where a
clear excess of power in the solar direction is seen,
consistent with the expectation~({\small{\cite{ped}}}).
This conclusion is encouraging for ANITA since it shows that the
measured RF noise is consistent with expected noise levels.

\paragraph{Impulse Noise Analysis Results}
ANITA-lite measured about 130,000 distinct events, with an
approximate rate of 5 events per minute. 
Fig.~\ref{fig6} shows examples of the major classes of impulsive
noise encountered.
The vast majority of triggered events are due to local payload
interference such as switching noise from the Support Instrument
Package.
The duration of this class of events exceeds several hundred
nanoseconds and a few events exhibit circular polarization for a
portion of the pulse. The bottom two panes show an example of a
synthetic pulse injected
into the data stream to simulate a true signal event. The pulse is 
coherent and aligned across all channels.
Preliminary studies of event selection procedures were conducted. They
yield no passing events, while $97\%$ of simulated Askaryan-induced
impulses with amplitude $5\sigma$ above noise survive~({\small{\cite{ped}}}).

\section{Attenuation Studies of Antarctic Ice}
The attenuation length of the ice beneath the South Pole Station was
measured in 2004~({\small{\cite{Steve_04}}}). 
A pair of broadband Ultra-High Frequency (UHF) horns, with range
from 200 to 700 MHz, were used to make echograms by reflecting radar
pulses off the bottom of the ice cap (depth 2810 m), and measuring the
return amplitude in a separate receiver.
Antarctic ice exhibits a horizontally layered structure, which creates
small variations in the index of refraction. The near
vertical penetration of the radio signal through the ice strata
minimizes the impact of reflections due to variation of the index
refraction.
By making the assumption that the reflection
coefficient off the bottom is unity, one can derive a lower limit on
the attenuation length. Because of the
logarithmic dependence of the derived attenuation length on most of
the parameters, the uncertainty in the attenuation length is
relatively small. A summary of the results is shown in
Fig.~\ref{fig7}, where the error bars are an estimate of the 2$\sigma$
systematic errors such as uncertainty in temperature profile of the
ice, reflection coefficient, and antenna responses.
The radio pulse propagates through ice which varies from $-50^{\circ}$C
at the surface to near $0^{\circ}$C at the bottom. 
We correct the results to $-45^{\circ}$C, a typical temperature for Antarctic ice.
At this temperature the field attenuation length is
$\ge 1$~km between 200-700 MHz. These encouraging results define one
of the most fundamental physical properties necessary for the success
of ANITA.

\section{ANITA Sensitivity}
\begin{figure}[ht]
\includegraphics[width=2.2 truein]{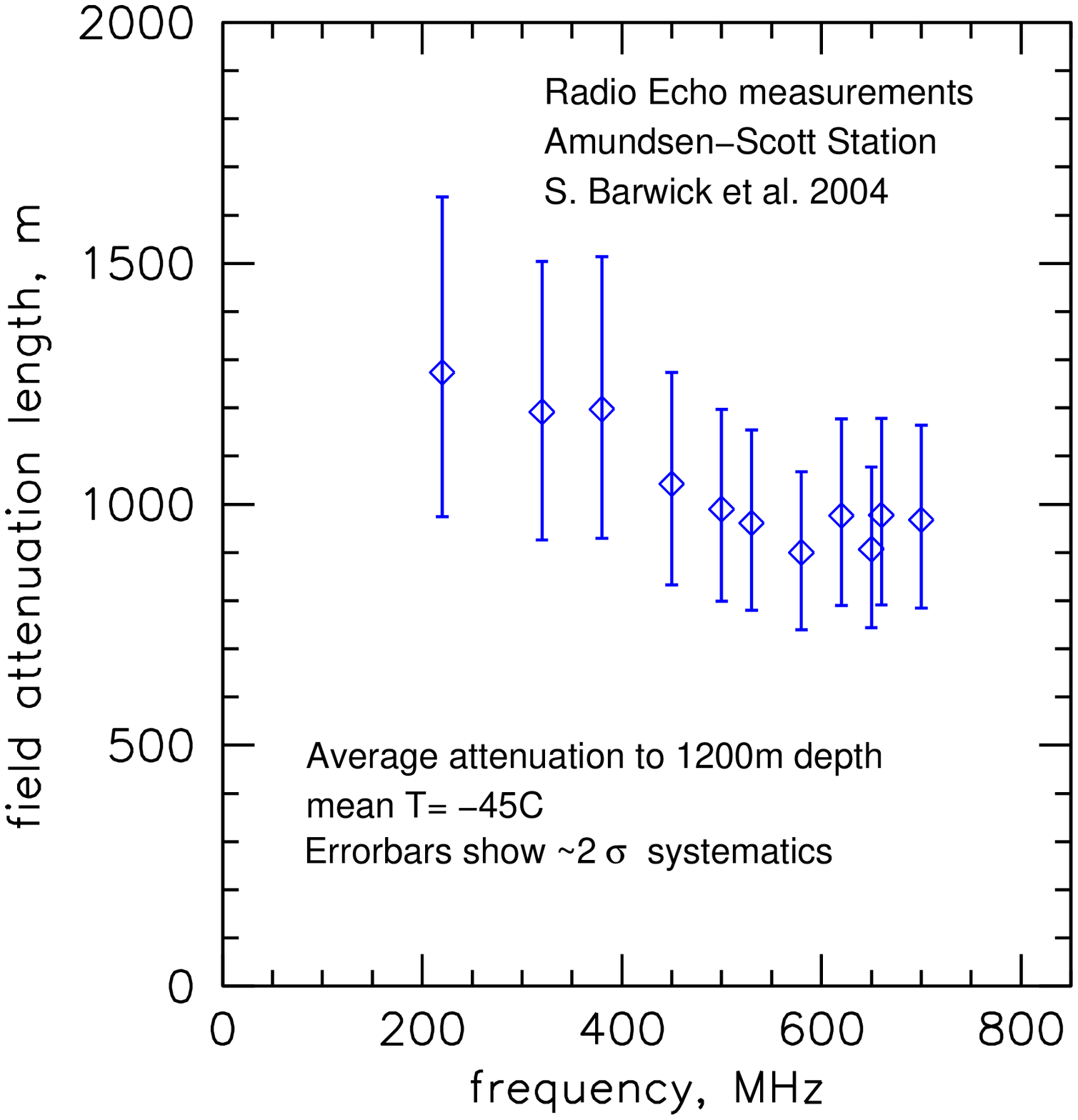}
~~~~~\includegraphics[width=2.2 truein]{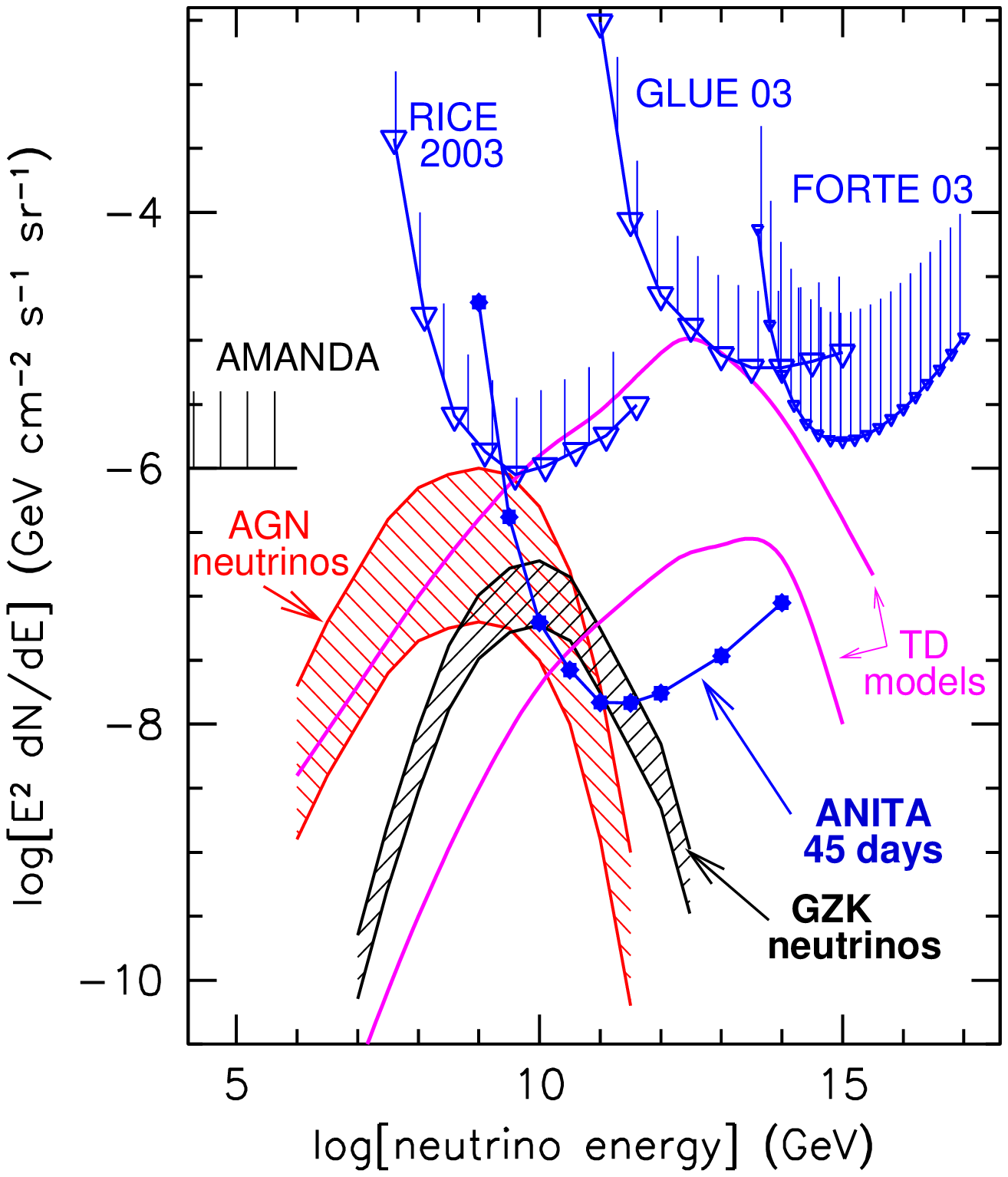}
\vskip-0.2in
\sidebyside
{\caption{\label{fig7} Field attenuation length versus radio frequency 
for 2004 measurements at the South Pole.}}
{\caption{\label{fig8} Neutrino models and experimental limits
    (90\% C.L.) along with a recent estimate single-event sensitivity
    per energy decade of ANITA for the baseline mission of 3 flights
    that achieve 45 days of total exposure, assuming that ice is
    observable from ANITA payload 67\% of the time.}}
\end{figure}

Fig.~\ref{fig8} shows various neutrino models, experimental limits,
and a recent (late 2004) estimate of the sensitivity of
ANITA for the baseline 3 LDB flights.
ANITA will achieve $\sim2$ orders of magnitude 
improvement over existing limits on neutrino fluxes in the relevant energy
regime. Assuming a 67\% exposure over deep ice, typical of an LDB flight, we
expect between 5-15 events from standard model GZK
fluxes~({\small{\cite{Engel:2001hd,Stanev02}}}), with the
uncertainty primarily arising from assumptions about cosmic ray source
evolution. Conversely, an observation of no events would reject {\em all}
standard GZK models at the $\ge$ 99\% C.L.
We stress that the GZK neutrino fluxes shown are
predictions of great importance to both high energy physics and our
understanding of cosmic-ray production and propagation. 
A non-detection of these fluxes by ANITA would suggest non-standard
phenomena in either particle physics or the astrophysics of cosmic ray
propagation.

\section{Conclusion}
ANITA will use radio emission from the cascade induced
by a neutrino interaction within the Antarctic
ice cap to detect UHE neutrino interactions that occur within a million
square km area.
The remarkable transparency of Antarctic ice to radio waves makes this
experiment possible, and the enormous volume of ice that can be
simultaneously monitored leads to an unparalleled sensitivity to
neutrinos in the energy range of 0.1 to 100 EeV.
Three separate flights provide 45 days of livetime, which can be
increased by multiple orbits around the South Pole. It is possible,
under optimal conditions and the use of multiple orbits around
Antarctica, to obtain an integrated livetime approaching 100 days.
In December 2001 the TIGER instrument obtained a record 31.8 day
flight by circling twice around the Antarctic continent.
Even the first ANITA flight will constrain model predictions of GZK
neutrino fluxes and provide insight to particle physics at the energy
frontier.

\begin{acknowledgments}
{\small
This research was supported by the following agencies: NASA Research
Opportunities in Space Science (ROSS) -- The UH grant number is NASA
NAG5-5387, Research Opportunities in Space Science -- NSF
and Raytheon Polar Services for Antarctic Support -- 
NSBF for Balloon Launching and Operations -- TIGER Collaboration for allowing
ANITA-lite to fly as a piggyback -- NATO Advanced Study Institute for
providing the Full Scholarship at Erice.}
\end{acknowledgments}

\end{document}